\documentclass[aps,showkeys,twocolumn,preprintnumbers,showpacs,amsmath,amssymb,superscriptaddress,floatfix,nofootinbib]{revtex4}
\usepackage{graphicx,color,dcolumn,booktabs,bm}
\usepackage{longtable,lscape}
\usepackage{txfonts}
\usepackage{overpic}
\usepackage{epsfig}
\usepackage{amssymb}
\usepackage{rotating}
\usepackage{epstopdf}
\usepackage{appendix}
\usepackage{indentfirst}
\usepackage{feynmf}   
\usepackage{slashed}  
\usepackage{cases}
\usepackage{color}
\usepackage{multirow}
\usepackage{graphicx,color,dcolumn,booktabs,bm}
\usepackage{cases}
\usepackage{array}

\graphicspath{{Figures/}} %

\usepackage[colorlinks, citecolor=blue,anchorcolor=red,menucolor=red, linkcolor=red,filecolor=red,runcolor=red,urlcolor=blue,frenchlinks=red]{hyperref}

\begin{document}
\title{Strong decays of the low-lying doubly bottom baryons }
\author{Hui-Zhen He}
\affiliation{  Department
of Physics, Hunan Normal University,  Changsha 410081, China }

\affiliation{ Synergetic Innovation
Center for Quantum Effects and Applications (SICQEA), Changsha 410081,China}

\affiliation{  Key Laboratory of
Low-Dimensional Quantum Structures and Quantum Control of Ministry
of Education, Changsha 410081, China}

\author{Wei Liang}
\affiliation{  Department
of Physics, Hunan Normal University,  Changsha 410081, China }

\affiliation{ Synergetic Innovation
Center for Quantum Effects and Applications (SICQEA), Changsha 410081,China}

\affiliation{  Key Laboratory of
Low-Dimensional Quantum Structures and Quantum Control of Ministry
of Education, Changsha 410081, China}

\author{Qi-Fang L\"u \footnote{Corresponding author} } \email{lvqifang@hunnu.edu.cn} %
\affiliation{  Department
of Physics, Hunan Normal University,  Changsha 410081, China }

\affiliation{ Synergetic Innovation
Center for Quantum Effects and Applications (SICQEA), Changsha 410081,China}

\affiliation{  Key Laboratory of
Low-Dimensional Quantum Structures and Quantum Control of Ministry
of Education, Changsha 410081, China}

\begin{abstract}
In this work, we adopt the $^3P_0$ model to investigate the strong decays of the low-lying doubly bottom baryons in the  $j-j$ coupling scheme systematically. In this scheme, we construct the formulism of $^3P_0$ model under the spectator assumption, and then the heavy diquark symmetry is preserved automatically. Our results show that some of the $\lambda$-mode $\Xi_{bb}(1P)$ and $\Omega_{bb}(1P)$ states are narrow, which have good potentials to be observed by future experiments. For the low-lying $\rho$-mode and $\rho$-$\lambda$ hybrid states, the Okubo-Zweig-Iizuka-allowed strong decays are highly suppressed and they should be extremely narrow. Future experiments can test our phenomenological predictions at the quark level.

\end{abstract}


\maketitle

\section{Introduction}
In recent years, significant progresses were achieved in heavy baryon spectroscopy both experimentally and theoretically, and the searching for new heavy baryons has become one of intriguing topics in hadron physics. The investigations on heavy baryons can help us to establish their mass spectra, enrich the structures of conventional hadrons, and deepen the understandings of heavy quark symmetry. The heavy baryons can be simply divided into three categories according to the number of heavy quarks: singly, doubly, and fully ones. For the singly heavy baryons, a series of structures have been observed experimentally, which can be interpreted as the conventional baryons  reasonably  within various theoretical works. Compared with the singly heavy ones, the experimental signals for doubly and fully heavy baryons are scare, and more attentions should be paid to hunt for these new structures.  

In 2002, the SELEX Collaboration announced some evidence of a doubly charmed baryon $\Xi_{cc}^{+}$ with a mass of 3519 $\pm$ 1 MeV in the $\Lambda_c^+K^-\pi^+$ final state~\cite{Mattson:2002vu}, and also observed it in  the $pD^+K^-$ mode subsequently~\cite{Ocherashvili:2004hi}. However, this structure was not confirmed by the following experiments of FOCUS, BaBar, Belle, and LHCb Collaborations~\cite{Ratti:2003ez,Aubert:2006qw,Chistov:2006zj,Aaij:2013voa}. In 2017, the LHCb Collaboration discovered a highly significant structure $\Xi_{cc}^{++}(3621)$ in the $\Lambda_{c}^{+}K^{-}\pi^{+}\pi^{+}$ mass spectrum, which has a mass of 3621.40 $\pm$ 0.72 $\pm$ 0.27 $\pm$ 0.14 MeV~\cite{Aaij:2017ueg}. Then, the LHCb Collaboration further investigated the precise mass, lifetime, and decay channels of $\Xi_{cc}^{++}(3621)$~\cite{Aaij:2018gfl,Aaij:2018wzf,Aaij:2019dsx,Aaij:2019zxa,Aaij:2019uaz}. Furthermore, the LHCb Collaboration tried to find out more doubly heavy baryons, but no extra signal has been observed so far~\cite{Aaij:2019jfq,Aaij:2020vid,Aaij:2021uym,Aaij:2021doa}. The searches goes on, and fruitful heavy baryon spectroscopy are waiting for experimentalists to explore and discover.
 
Before the observation of $\Xi_{cc}^{++}(3621)$, there were lots of works to predict the mass spectra of doubly heavy baryons within various methods, such as potential models~\cite{Kiselev:2001fw,Ebert:1996ec,Ebert:2002ig,Gershtein:2000nx,Roberts:2007ni,Giannuzzi:2009gh,Martynenko:2007je,Valcarce:2008dr,Eakins:2012jk,Shah:2017liu}, heavy quark symmetry and mass formulas~\cite{Roncaglia:1995az,Cohen:2006jg,Karliner:2014gca}, Regge behaviors~\cite{Wei:2015gsa,Wei:2016jyk}, QCD sum rule~\cite{Zhang:2008rt,Tang:2011fv,Wang:2010hs,Aliev:2012ru,Aliev:2012nn,Aliev:2012iv}, lattice QCD~\cite{Liu:2009jc,Brown:2014ena,Padmanath:2015jea} and so on. Besides the mass spectra, the weak and radiative decays of the doubly heavy baryons are also extensively discussed in the literature~\cite{Faessler:2001mr,Faessler:2009xn,Albertus:2009ww,White:1991hz,Li:2017ndo,Yu:2017zst,Ebert:2004ck,Roberts:2008wq,Branz:2010pq,Hackman:1977am,Bernotas:2013eia,Dai:2000hza,Albertus:2010hi}, which provide helpful information for the experimental searches. The observation of $\Xi_{cc}^{++}(3621)$ interested many theorists immediately, and plenty of works have been done in recent years to discuss their mass spectra~\cite{Kiselev:2017eic,Chen:2017sbg,Lu:2017meb,Wang:2017qvg,Weng:2018mmf,Karliner:2018hos,Richard:2018yrm,Yu:2018com,Li:2019ekr,Faustov:2020gun,Braaten:2020nwp}, weak and radiative decays~\cite{Wang:2017mqp,Gutsche:2017hux,Sharma:2017txj,Shi:2017dto,Cui:2017udv,Gershon:2018gda,Zhang:2018llc,Ridgway:2019zks,Cheng:2019sxr,Gerasimov:2019jwp,Ke:2019lcf,Shi:2020qde,Pan:2020qqo,Han:2021azw,Li:2021rfj}, strong decays~\cite{Xiao:2017udy,Mehen:2017nrh,Ma:2017nik,Xiao:2017dly,Yan:2018zdt}, magnetic moments~\cite{Li:2017cfz,Meng:2017dni,Ozdem:2018uue,Shi:2021kmm}, productions~\cite{Yao:2018zze,Niu:2018ycb,Li:2020ggh,Berezhnoy:2018krl,Berezhnoy:2020aox}, and other relevant topics~\cite{Yao:2018ifh,Meng:2018zbl,Mehen:2019cxn,Olamaei:2020bvw,Soto:2020pfa,Luchinsky:2020fdf,Andreev:2020xor}. 
 
Among these extensive theoretical works, there were only a few studies on strong decay behaviors. At the hadronic level, several works investigated the strong decays of low-lying doubly heavy baryons within the effective Lagrangian approach by considering the heavy diquark symmetry~\cite{Hu:2005gf,Ma:2015cfa,Mehen:2017nrh,Ma:2017nik,Yan:2018zdt}. At the quark level, the authors adopted the chiral quark model and $^3P_0$ model to investigate doubly heavy baryons, and found that the strong decays for low-lying $\rho$-mode states are highly suppressed~\cite{Eakins:2012fq,Xiao:2017udy,Xiao:2017dly}. It can be seen that the predictions from different models and parameters are not consistent with each other, or even differ vastly. Especially, these previous works mainly focus on the strong decays of doubly charmed ones, while the studies on doubly bottom baryons are scarce. 
 
Actually, the investigations on doubly bottom baryons may be easier compared with the doubly charmed ones. With the  heavy bottom quark mass, the two bottom quarks stay close to each other like a static color source,  and the light quark is shared by these two bottom quarks and circles around this source~\cite{Savage:1990di}. In this picture, a doubly bottom baryon seems to be a bottom or bottom-strange meson, where the $bb$ subsystem looks like a antibottom quark. It is hoped that the doubly bottom baryons and bottom/bottom-strange mesons may have similar decay patterns, and then the heavy diquark symmetry emerges and should be preserved approximately.      
 
In this work, we perform a systematic analysis of strong decays for the low-lying doubly bottom states within the $^3P_0$ model. In Ref.~\cite{Xiao:2017dly}, the authors adopted the $L-S$ coupling scheme to study the properties of doubly charmed baryons, since the heavy diquark symmetry may be broken significantly in the charmed sector. Here, we use the $j-j$ coupling scheme to calculate the strong decays of doubly bottom baryons owing to the rather heavy mass of bottom quark. In this scheme, both the light quark spin and heavy diquark spin can be treated as good quantum numbers approximately. Our results indicate that some of $\lambda$-mode $\Xi_{bb}(1P)$ and $\Omega_{bb}(1P)$ states are narrow, which have good potentials to be observed by future experiments. However, the strong decays of the low-lying $\rho$-mode and $\rho$-$\lambda$ hybrid states are highly suppressed owing to the orthogonality of spatial wave functions. Moreover, the masses and strong decay behaviors of doubly bottom baryons show similar patterns to the bottom/bottom-strange mesons, which suggests that the diquark correlation should play an essential role in these systems. 

This paper is organized as follows. The framework of strong decays for doubly bottom baryons are introduced in Sec~\ref{model}. We present the numerical results and discussions for the low-lying doubly bottom baryons in Sec~\ref{low-lying}. A summary is presented in the last section.

\section{Model}{\label{model}}

\subsection{Coupling scheme}

\begin{figure}[!htbp]
	\includegraphics[scale=0.55]{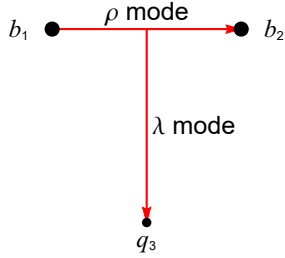}
	\vspace{0.0cm} \caption{The Jacobi coordinates of doubly bottom baryons. The $b_1$ and $b_2$ stand for bottom quarks, and $q_3$ corresponds to a light quark. The $\boldsymbol \rho = (\boldsymbol{r_1}-\boldsymbol{r_2})/\sqrt{2}$ is the relative coordinate between two bottom quarks, and the $\boldsymbol \lambda = (\boldsymbol{r_1}+\boldsymbol{r_2}-2\boldsymbol{r_3})/\sqrt{6}$ is the relative coordinate between the light quark and bottom quark subsystem.}\label{jacobi}
\end{figure}

To calculate the strong decays of doubly heavy baryons, one should label the low-lying states with certain quantum numbers in a coupling scheme firstly. The Jacobi coordinates of doubly bottom baryons are presented in Figure~\ref{jacobi}. With this definition, we can classify the excitations of doubly bottom baryons into three types: $\rho$-mode, $\lambda$-mode, and $\rho$-$\lambda$ hybrid excitations.     

With the Jacobi coordinates, we can use a series of quantum numbers to characterize the excitations of spatial wave functions. Here, the $n_\rho$ and $l_\rho$ are the radial and orbital quantum numbers between the two bottom
quarks, respectively; similarly, the $n_\lambda$ and $l_\lambda$ correspond to the radial and orbital quantum numbers between the light quark and bottom quark subsystem, respectively. The $N=2n_\lambda+2n_\rho+l_\rho+l_\lambda$ is the principle quantum number, and the low-lying doubly bottom baryons denote the states up to $N=2$ shell. The $S_\rho$ and $J_\rho$ stands for the total spin and total angular momentum of the two bottom quarks, respectively. The $\boldsymbol j = \boldsymbol{l_\lambda} +\boldsymbol{s_3}$ represents the light quark spin, and $J^P$ is the spin-parity of a doubly bottom baryon. Then, the coupling bases can be expressed as 
\begin{eqnarray}
	\vert{J^{P}}, j\rangle=
	\Big \vert \left [(l_{\rho}S_{\rho})_{J_\rho}(l_{\lambda}s_{3})_j \right ] _{J^P} \Big \rangle.
\end{eqnarray}
In the heavy quark limit, the heavy diquark spin $J_{\rho}$ and the light quark spin $j$ should be preserved well. With finite bottom quark masses, the states with same $J^P$ can be mixed with each other. However, the physical states should be much closer to these coupling bases to preserve the heavy diquark symmetry approximately, especially for the doubly bottom systems. Hence, we adopt these coupling bases to investigate the strong decay behaviors for doubly bottom baryons in present work. 

\begin{figure*}[!htbp]
	\includegraphics[scale=0.8]{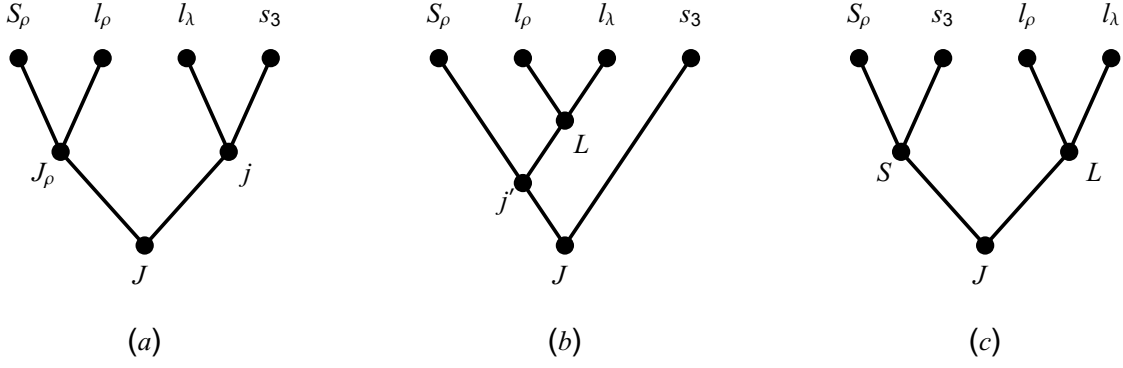}
	\vspace{0.0cm} \caption{Three coupling scheme in tree representations.}\label{scheme}
\end{figure*}

The above coupling scheme differs from the $j-j$ coupling adopted in the singly bottom baryons, because the light quark spins for the singly and doubly heavy systems are quite different. In singly heavy baryons, the $\boldsymbol \rho$ is the relative coordinate between two light quarks, and the $\boldsymbol \lambda$ is the relative coordinate between the heavy quark and light quark subsystem. For singly heavy baryons, the corresponding coupling basis that preserving the light quark spin $ j^\prime$ can be written as
\begin{eqnarray}
	\vert{J^{P}}, j^\prime\rangle=
	\Bigg \vert\left\{ \left[(l_{\rho}l_{\lambda})_{L}S_{\rho}\right]_{j^\prime}s_{3}\right\} _{J^P} \Bigg \rangle,
\end{eqnarray}
where the $L$ and $ j^\prime$ represents the quantum numbers of total orbital angular momentum and light quark spin, respectively. It can be noticed that there is a linear relationship between these two $j-j$ coupling bases
\begin{eqnarray}
\vert{J^{P}}, j\rangle & =& \sum_{L ~j^\prime} (-1)^{S_{\rho}+L+s_{3}+J+2J_{\rho}+2l_{\lambda}}\nonumber \\ && \times
\sqrt{(2j+1)(2j^\prime +1)(2J_{\rho}+1)(2L+1)} \nonumber \\ && \times \left\{ \begin{array}{ccc}
		 J_{\rho}& S_{\rho} & j^\prime\\
		s_{3} & J & j
	\end{array}\right\} 
	\left\{ \begin{array}{ccc}
		 l_{\rho}& S_{\rho} & J_{\rho}\\
		J & l_{\lambda} & L
	\end{array}\right\}\vert{J^{P}}, j^\prime\rangle.
\end{eqnarray}
Furthermore, the $L-S$ coupling scheme is also commonly used in the literature, and the coupling bases are
\begin{eqnarray}
	\vert ^{2S+1}L_{J}\rangle=
	\Bigg \vert\left[ (l_{\rho}l_{\lambda})_{L}(S_{\rho}s_{3})_{S} \right]_{J^P} \Bigg \rangle.
\end{eqnarray}
Then the relations among the $j-j$ and $L-S$ coupling scheme can be written as
\begin{eqnarray}
\vert{J^{P}}, j^\prime\rangle & =&  \sum_{S} (-1)^{S_{\rho}+L+s_{3}+J}\sqrt{(2j^\prime +1)(2S+1)} \nonumber \\ && \times \left\{ \begin{array}{ccc}
		L & J & j^\prime\\
		s_{3} & S_{\rho} & S
	\end{array}\right\} \vert ^{2S+1}L_{J}\rangle.
\end{eqnarray}
and 
\begin{eqnarray}
\vert{J^{P}}, j\rangle & =& \sum_{L~S} \sqrt{(2j+1)(2S+1)(2L+1)(2J_{\rho}+1)} \nonumber \\ && \times \left\{ \begin{array}{ccc}
		l_{\rho} & S_{\rho} & J_{\rho}\\
		 l_{\lambda}   & s_{3}& j\\
		 L&S&J\\
	\end{array}\right\} \vert ^{2S+1}L_{J}\rangle.
\end{eqnarray}
The tree representations for three different coupling schemes are shown in Figure~\ref{scheme} for reference. 

It should be mentioned that the a physical state can be the mixture of theoretical states with same $J^{P}$, and the final results for strong decays are independent with different coupling schemes. However, because of the heavy diquark symmetry, doubly bottom baryons favor the $\vert{J^{P}}, j\rangle$ bases, that is, the mixing angles between physical and theoretical states $\vert{J^{P}}, j\rangle$ should be small and the mixing effects can be ignored here. Moreover, we employ the notations $\Xi_{bb}(J^P, j)$ and $\Omega_{bb}(J^P, j)$ to represent the orbital excited states $\vert{J^{P}}, j\rangle$ in present work.  

\subsection{Mass}
For the masses of low-lying $\Xi_{bb}$ and $\Omega_{bb}$ states, we adopt the theoretical predictions from the relativistic quark model~\cite{Ebert:2002ig}. The masses of $\lambda$-mode S-wave and P-wave doubly bottom baryons are listed in Table~\ref{Mass}. Here, we treat the light quark spin $j$ as a good quantum number and neglect the small mixtures among the $P$-wave states. Also, the masses of low-lying $\rho$-mode states are not listed, as the their strong decays are highly suppressed under the spectator assumption of $^3P_0$ model. For the $\lambda$-mode $\Xi_{bb}(1D)$ and $\Omega_{bb}(1D)$ states, the authors did not perform their masses within the relativistic quark model. Fortunately, we can estimate their average masses with the help of Regge trajectory and heavy diquark symmetry.      

\begin{figure}[!htbp]
	\includegraphics[scale=0.9]{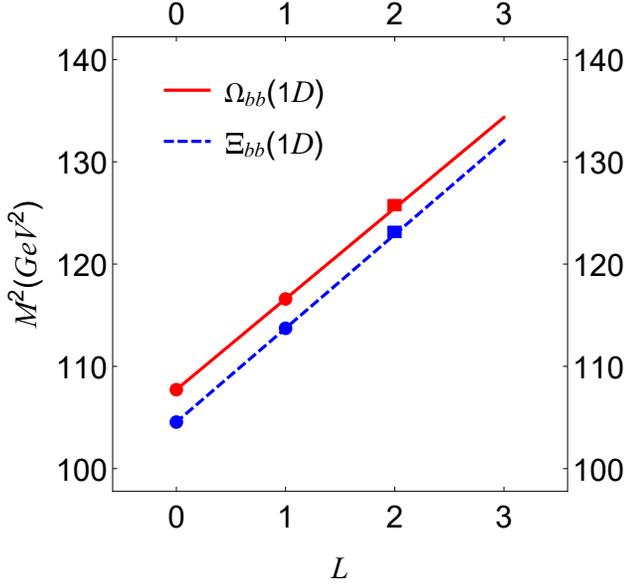}
	\vspace{0.0cm} \caption{The Regge trajectories of $\Xi_{bb}(1D)$ and $\Omega_{bb}(1D)$ states.}\label{regge}
\end{figure}

From Table~\ref{Mass}, the average masses of the $\Xi_{bb}(1S)$, $\Xi_{bb}(1P)$, $\Omega_{bb}(1S)$ and $\Omega_{bb}(1P)$ states are
\begin{eqnarray}
	\bar m[\Xi_{bb}(1S)] = \lbrace 2m[\Xi_{bb}]+4m[\Xi_{bb}^*]\rbrace /6 = 10225~\rm{MeV},
\end{eqnarray}
\begin{eqnarray}
\bar m[\Xi_{bb}(1P)] = 10664~\rm{MeV},
\end{eqnarray}
\begin{eqnarray}
	\bar m[\Omega_{bb}(1S)] = \lbrace 2m[\Omega_{bb}]+4m[\Omega_{bb}^*]\rbrace /6 =10379~\rm{MeV},
\end{eqnarray}
\begin{eqnarray}
	\bar m[\Omega_{bb}(1P)] = 10798~\rm{MeV}.
\end{eqnarray}
Their Regge trajectories of $\Xi_{bb}(1D)$ and $\Omega_{bb}(1D)$ between mass and orbital angular momentum are plotted in Figure.~\ref{regge}. From these trajectories, we can obtain the estimated average masses of $\Xi_{bb}(1D)$ and $\Omega_{bb}(1D)$ states are about 11086 MeV and 11201 MeV, respectively. 

Meanwhile, because of the doubly diquark symmetry, the mass gaps of doubly bottom baryons should be similar to that of bottom/bottom-strange mesons. With the same relativistic quark model, the mass gaps of bottom/bottom-strange mesons are estimated as~\cite{Ebert:2009ua} 
\begin{eqnarray}
	\bar m[B(1P)] - \bar m[B(1S)] = 430~\rm{MeV},
\end{eqnarray}
\begin{eqnarray}
	\bar m[B(1D)] - \bar m[B(1S)] = 791~\rm{MeV},
\end{eqnarray}
\begin{eqnarray}
	\bar m[B_s(1P)] - \bar m[B_s(1S)] = 440~\rm{MeV},
\end{eqnarray}
\begin{eqnarray}
	\bar m[B_s(1D)] - \bar m[B_s(1S)] = 796~\rm{MeV}.
\end{eqnarray}
It can be seen that the mass gaps 
$\bar m[\Xi_{bb}(1P)] - \bar m[\Xi_{bb}(1S)] \simeq \bar m[B(1P)] - \bar m[B(1S)]$ and $\bar m[\Omega_{bb}(1P)] - \bar m[\Omega_{bb}(1S)] \simeq \bar m[B_s(1P)] - \bar m[B_s(1S)]$ preserve the heavy diquark symmetry approximately. Hence, we have 
\begin{eqnarray}
\bar m[\Xi_{bb}(1D)] - \bar m[\Xi_{bb}(1S)] = 791 ~\rm{MeV},
\end{eqnarray}
\begin{eqnarray}
	\bar m[\Omega_{bb}(1D)] - \bar m[\Omega_{bb}(1S)] = 796~\rm{MeV}.
\end{eqnarray}  
Then, the predicted average masses of $\Xi_{bb}(1D)$ and $\Omega_{bb}(1D)$ states from the heavy diquark symmetry are about 11016 MeV and 11175 MeV, respectively. Also, the approximate SU(3) flavor symmetry for light quarks is preserved well.  

With the above two approaches, we obtain the mass ranges for $\Xi_{bb}(1D)$ and $\Omega_{bb}(1D)$ states, which are listed in Table~\ref{Mass}. Due to the small mass splittings, the estimated average masses of $\Xi_{bb}(1D)$ and $\Omega_{bb}(1D)$ states are enough to investigate their strong decays.

\begin{table}[htb]
	\begin{center}
		\caption{ \label{Mass} Notations, quantum numbers, and masses of the doubly bottom baryons. The values are in MeV. }
		\renewcommand{\arraystretch}{1.5}
		\begin{tabular*}{8.6cm}{@{\extracolsep{\fill}}p{1.3cm}<{\centering}p{0.4cm}<{\centering}p{0.4cm}<{\centering}p{0.4cm}<{\centering}p{0.4cm}<{\centering}p{0.4cm}<{\centering}p{0.4cm}<{\centering}p{0.4cm}<{\centering}p{0.4cm}<{\centering}p{1.8cm}<{\centering}}
			\hline\hline
			State     	& $n_{\rho}$ & $n_{\lambda}$ & $l_{\rho}$ & $l_{\lambda}$ & $S_{\rho}$ & $J_{\rho}$ &   $j$	&  $J^P$&Mass\\
			\hline
$\Xi_{bb} (1S)$	 	&	0	&	0	&	0	&	0	&	1       &1 &	$\frac{1}{2}$	&	$\frac{1}{2}^+$     & 10202 \\
$\Xi_{bb}^{*}(1S)$		&	0	&	0	&	0	&	0	&	1    &1      &	$\frac{1}{2}$	&	$\frac{3}{2}^+$   &	10237  \\ 
$\Xi_{bb} (2S)$	 	&	0	&	1	&	0	&	0	&	1       &1 &	$\frac{1}{2}$	&	$\frac{1}{2}^+$     & 10832  \\
$\Xi_{bb}^{*}(2S)$		&	0	&	1	&	0	&	0	&	1    &1      &	$\frac{1}{2}$	&	$\frac{3}{2}^+$   &	10860  \\
$\Xi_{bb} (\frac{1}{2}^-, \frac{1}{2})$   &	0	&	0	&	0	&	1	&	1   &1 &	$\frac{1}{2}$	&   $\frac{1}{2}^-$ &	10675 \\
$\Xi_{bb} (\frac{3}{2}^-, \frac{1}{2}) $  		&	0	&	0	&	0	&	1	&		1&1   &	$\frac{1}{2}$	&   $\frac{3}{2}^-$  &	10694 \\
$\Xi_{bb} (\frac{1}{2}^-,\frac{3}{2})$	   &		0	&	0	&	0	&	1	&	1& 1	  &	$\frac{3}{2}$	&	$\frac{1}{2}^-$    &	10632 \\
$\Xi_{bb} (\frac{3}{2}^-, \frac{3}{2})$  		&	0	&	0	&	0	&	1	&	1 &	1  &	$\frac{3}{2}$	&   $\frac{3}{2}^-$    &	10647 \\
$\Xi_{bb} (\frac{5}{2}^-, \frac{3}{2})$	  		&	0	&	0	&	0	&	1	&	1  &	1 &	$\frac{3}{2}$	&	$\frac{5}{2}^-$   &	10661 \\
$\Xi_{bb} (\frac{1}{2}^+, \frac{3}{2})$   &	0	&	0	&	0	&	2	&	1  &1   &	$\frac{3}{2}$	&   $\frac{1}{2}^+$ &	11016-11086 \\
$\Xi_{bb} (\frac{3}{2}^+, \frac{3}{2}) $   	&	0	&	0	&	0	&	2	&	1   &1&	$\frac{3}{2}$	&   $\frac{3}{2}^+$ &	11016-11086 \\
$\Xi_{bb} (\frac{5}{2}^+, \frac{3}{2})$	              	&	0	&	0	&	0	&	2	&	1 &1  &	$\frac{3}{2}$	&	$\frac{5}{2}^+$  &	11016-11086\\
$\Xi_{bb} (\frac{3}{2}^+, \frac{5}{2})$           		&	0	&	0	&	0	&	2	&	1  &1 &	$\frac{5}{2}$	&   $\frac{3}{2}^+$ &	11016-11086\\
$\Xi_{bb} (\frac{5}{2}^+, \frac{5}{2})$	              		&	0	&	0	&	0	&	2	&	1  &1 &	$\frac{5}{2}$	&	$\frac{5}{2}^+$  &	11016-11086\\
$\Xi_{bb} (\frac{7}{2}^+, \frac{5}{2})$	              &	0	&	0	&	0	&	2	&	1  &1&	$\frac{5}{2}$	&	$\frac{7}{2}^+$   &	11016-11086\\		
$\Omega_{bb} (1S)$	 &	0	&	0	&	0	&	0	&	1  &1 &	$\frac{1}{2}$	& $\frac{1}{2}^+$  & 10359 \\
$\Omega_{bb}^{*} (1S)$	&	0	&	0	&	0	&	0	&	1  &1 &	$\frac{1}{2}$	& $\frac{3}{2}^+$   &10389  \\
$\Omega_{bb} (2S)$	 &	0	&	1	&	0	&	0	&	1  &1 &	$\frac{1}{2}$	& $\frac{1}{2}^+$  & 10970  \\
$\Omega_{bb}^{*} (2S)$	&	0	&	1	&	0	&	0	&	1  &1 &	$\frac{1}{2}$	& $\frac{3}{2}^+$   &10992  \\
$\Omega_{bb} (\frac{1}{2}^-, \frac{1}{2})$   	&	0	&	0	&	0	&	1	&	1 &1  &	$\frac{1}{2}$	& $\frac{1}{2}^-$  &10804 \\
$\Omega_{bb} (\frac{3}{2}^-, \frac{1}{2}) $  	&	0	&	0	&	0	&	1	&	1  &1&	$\frac{1}{2}$	& $\frac{3}{2}^-$    &10821 \\
$\Omega_{bb} (\frac{1}{2}^-, \frac{3}{2})$	  	&	0	&	0	&	0	&	1	&	1   &1 &	$\frac{3}{2}$	& $\frac{1}{2}^-$  &10771 \\
$\Omega_{bb} (\frac{3}{2}^-, \frac{3}{2})$   	&	0	&	0	&	0	&	1	&	1 &1  &	$\frac{3}{2}$	& $\frac{3}{2}^-$  &10785 \\
$\Omega_{bb} (\frac{5}{2}^-, \frac{3}{2})$		&	0	&	0	&	0	&	1	&	1 &1   &	$\frac{3}{2}$	& $\frac{5}{2}^-$  &10798 \\
$\Omega_{bb} (\frac{1}{2}^+, \frac{3}{2})$                		&	0	&	0	&	0	&	2	&	1  &	1&$\frac{3}{2}$	& $\frac{1}{2}^+$   &11175-11201 \\
			$\Omega_{bb} (\frac{3}{2}^+, \frac{3}{2})$           	&	0	&	0	&	0	&	2	&	1 &1   &	$\frac{3}{2}$	& $\frac{3}{2}^+$ &11175-11201 \\
			$\Omega_{bb} (\frac{5}{2}^+, \frac{3}{2})$	           		&	0	&	0	&	0	&	2	&	1  &1  &	$\frac{3}{2}$	& $\frac{5}{2}^+$ &11175-11201\\
			$\Omega_{bb} (\frac{3}{2}^+, \frac{5}{2})$           	&	0	&	0	&	0	&	2	&	1   &1&	$\frac{5}{2}$	& $\frac{3}{2}^+$  &11175-11201 \\
			$\Omega_{bb} (\frac{5}{2}^+, \frac{5}{2})$	           		&	0	&	0	&	0	&	2	&	1  &1 &	$\frac{5}{2}$	& $\frac{5}{2}^+$  &11175-11201\\
			$\Omega_{bb} (\frac{7}{2}^+, \frac{5}{2})$	           	&	0	&	0	&	0	&	2	&	1  &1 &	$\frac{5}{2}$	& $\frac{7}{2}^+$  &11175-11201\\
			\hline\hline
		\end{tabular*}
	\end{center}
\end{table}

\subsection{$^3P_0$ Model}
In this work, we adopt the $^3P_0$ model to calculate the Okubo-Zweig-Iizuka-allowed two-body strong decays of the low-lying $\Xi_{bb}$ and $\Omega_{bb}$ states. In this model, a quark-antiquark pair with the quantum number $J^{PC}$ =$0^{++}$ is created from the vacuum and then groups into the final states ~\cite{Micu:1968mk}. This model has been employed to study the strong decays for different kinds of hadron systems with considerable successes~\cite{3p0model1,LeYaouanc:1977gm,3p0model2,3p0model4,3p0model5,3p0model6,Zhao:2016qmh,Chen:2007xf,Chen:2016iyi,Lu:2016bbk,Ferretti:2014xqa,Godfrey:2015dva,Segovia:2012cd,Lu:2020ivo,Liang:2020hbo,He:2021xrh}. In the nonrelativistic limit, to describe the decay process $A\rightarrow BC$, the transition operator $T$  in the $^3P_0$ model can be written as
\begin{eqnarray}
T&=&-3\gamma\sum_m\langle 1m1-m|00\rangle\int
d^3\boldsymbol{p}_4d^3\boldsymbol{p}_5\delta^3(\boldsymbol{p}_4+\boldsymbol{p}_5)\nonumber\\&&\times {\cal{Y}}^m_1\left(\frac{\boldsymbol{p}_4-\boldsymbol{p}_5}{2}\right)\chi^{45}_{1,-m}\phi^{45}_0\omega^{45}_0b^\dagger_{4i}(\boldsymbol{p}_4)d^\dagger_{4j}(\boldsymbol{p}_5),
\end{eqnarray}
where $\gamma$ is a dimensionless $q_4\bar{q}_5$ quark pair production strength, and $\boldsymbol{p}_4$ and $\boldsymbol{p}_5$ are the momenta of the created quark $q_4$ and antiquark  $\bar{q}_5$, respectively. The $i$ and $j$ are the color indices of the created quark and antiquark. $\phi^{45}_{0}=(u\bar u + d\bar d +s\bar s)/\sqrt{3}$, $\omega^{45}=\delta_{ij}$, and $\chi_{{1,-m}}^{45}$ are the flavor singlet, color singlet, and spin triplet wave functions of the  $q_4\bar{q}_5$, respectively. The ${\cal{Y}}^m_1(\boldsymbol{p})\equiv|p|Y^m_1(\theta_p, \phi_p)$ is the solid harmonic polynomial reflecting the $P$-wave momentum-space distribution of the created quark pair.

For an initial baryon $A$, the definition of the mock state is employed in present work, and it can be taken as
\begin{eqnarray}
&&|A(n^{2S_A+1}_AL_{A}\,\mbox{}_{J_A M_{J_A}})(\boldsymbol{P}_A)\rangle
\equiv \nonumber\\
&& \sqrt{2E_A}\sum_{M_{L_A},M_{S_A}}\langle L_A M_{L_A} S_A
M_{S_A}|J_A
M_{J_A}\rangle \int d^3\boldsymbol{p}_1d^3\boldsymbol{p}_2d^3\boldsymbol{p}_3\nonumber\\
&&\times \delta^3(\boldsymbol{p}_1+\boldsymbol{p}_2+\boldsymbol{p}_3-\boldsymbol{P}_A)\psi_{n_AL_AM_{L_A}}(\boldsymbol{p}_1,\boldsymbol{p}_2,\boldsymbol{p}_3)\chi^{123}_{S_AM_{S_A}}
\phi^{123}_A\omega^{123}_A\nonumber\\
&&\times  \left|q_1(\boldsymbol{p}_1)q_2(\boldsymbol{p}_2)q_3(\boldsymbol{p}_3)\right\rangle,
\end{eqnarray}
which satisfies the normalization condition
\begin{eqnarray}
\langle A(\boldsymbol{P}_A)|A(\boldsymbol{P}^\prime_A)\rangle=2E_A\delta^3(\boldsymbol{P}_A-\boldsymbol{P}^\prime_A).
\end{eqnarray}
The $\boldsymbol{p}_1$, $\boldsymbol{p}_2$, and $\boldsymbol{p}_3$ are the momenta of the quarks $q_1$, $q_2$, and $q_3$, respectively. $\boldsymbol{P}_A$ denotes the momentum of the initial state $A$.
$\chi^{123}_{S_AM_{S_A}}$, $\phi^{123}_A$, $\omega^{123}_A$, and
$\psi_{n_AL_AM_{L_A}}(\boldsymbol{p}_1,\boldsymbol{p}_2,\boldsymbol{p}_3)$ are the spin, flavor, color, and space wave functions of the baryon $A$. The definitions of the final baryon $B$ and meson $C$ are similar to the initial state $A$.

\begin{figure*}[!htbp]
\includegraphics[scale=0.8]{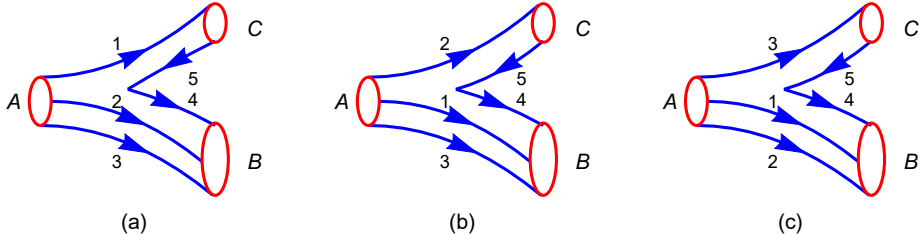}
\vspace{0.0cm} \caption{The baryon decay process $A\to B+C$ in the $^3P_0$ model.}
\label{qpc}
\end{figure*}

For the strong decays of doubly bottom baryons, there are three possible rearrangements,
\begin{eqnarray}
A(b_1,b_2,q_3)+P(q_4,\bar q_5)\to B(b_2,q_4,q_3)+C(b_1,\bar q_5),\\
A(b_1,b_2,q_3)+P(q_4,\bar q_5)\to B(b_1,q_4,q_3)+C(b_2,\bar q_5),\\
A(b_1,b_2,q_3)+P(q_4,\bar q_5)\to B(b_1,b_2,q_4)+C(q_3,\bar q_5).
\end{eqnarray}
These three possible rearrangements are shown in the Figure~\ref{qpc}. It can be seen that the first and second ones stand for the singly bottom baryon plus heavy meson channels, while the last one denotes the doubly bottom baryon plus light meson decay mode.  

With the transition operator $T$, the helicity amplitude ${\cal{M}}^{M_{J_A}M_{J_B}M_{J_C}}$ is defined as
\begin{eqnarray}
\langle
BC|T|A\rangle=\delta^3(\boldsymbol{P}_A-\boldsymbol{P}_B-\boldsymbol{P}_C){\cal{M}}^{M_{J_A}M_{J_B}M_{J_C}},
\end{eqnarray}
where the ${\cal{M}}^{M_{J_A}M_{J_B}M_{J_C}}$ is the helicity amplitude of a decay process $A\to B+C$. For the low-lying doubly bottom baryons, only the process $A(b_1,b_2,q_3)+\boldsymbol{P}(q_4,\bar q_5)\to B(b_1,b_2,q_4)+C(q_3,\bar q_5)$ is allowed owing to limited phase space. The helicity amplitude ${\cal{M}}^{M_{J_A}M_{J_B}M_{J_C}}$ can be expressed as,
\begin{eqnarray}
&&\delta^3(\boldsymbol{p}_B+\boldsymbol{p}_C-\boldsymbol{p}_A){\cal{M}}^{M_{J_A}M_{J_B}M_{J_C}} = \nonumber\\
&&- \gamma \sqrt{8E_AE_BE_C} \sum_{M_{J_{\rho_{A}}}, M_{j_A}, M_{l_{\rho_{A}}}, M_{S_{\rho_{A}}}, M_{l_{\lambda_{A}}}  } \nonumber\\
&& \sum_{M_{J_{\rho_{B}}}, M_{j_{B}}, M_{l_{\rho_{B}}}, M_{S_{\rho_{B}}}, M_{l_{\lambda_{B}}}}    \sum_{M_{L_C},M_{S_C}}  \sum_{M_{s_1},M_{s_2}, M_{s_3}, M_{s_4},M_{s_5}, m}\nonumber\\
&& \times  \langle J_{\rho_{A}} M_{J_{\rho_{A}}}j_{A} M_{j_{A}}|J_A M_{J_A}\rangle \langle l_{\rho_{A}} M_{l_{\rho_{A}}}S_{\rho_{A}}M_{S_{\rho_{A}}}|J_{\rho_{A}} M_{J_{\rho_{A}}} \rangle \nonumber\\ && \times \langle l_{\lambda_{A}}M_{l_{\lambda_{A}}}s_3M_{s_3}|j_{A} M_{j_{A}}\rangle \langle s_1M_{s_1}s_2M_{s_2}|S_{\rho_{A}}M_{S_{\rho_{A}}} \rangle \nonumber \\
&& \times \langle J_{\rho_{B}} M_{J_{\rho_{B}}}j_{B} M_{j_{B}}|J_B M_{J_B}\rangle \langle l_{\rho_{B}} M_{l_{\rho_{B}}}S_{\rho_{B}}M_{S_{\rho_{B}}}|J_{\rho_{B}} M_{j_B}\rangle \nonumber \\ 
&& \times \langle l_{\lambda_{B}}M_{l_{\lambda_{B}}}s_4M_{s_4}|j_{B} M_{j_{B}}\rangle \langle s_1M_{s_1}s_2M_{s_2}|S_{\rho_{B}}M_{S_{\rho_{B}}}\rangle\nonumber\\
&& \times \langle 1m 1-m|00\rangle \langle s_4M_{s_4}s_5M_{s_5}|1-m \rangle \nonumber\\
&& \times \langle L_C M_{L_C}S_CM_{S_C}|J_CM_{J_C}\rangle \langle s_3M_{s_3}s_5M_{s_5}|S_CM_{S_C}\rangle \nonumber\\
&& \times \langle \phi_B^{124} \phi_C^{35}|\phi_A^{123}\phi_0^{45}\rangle I^{M_{L_A}m}_{M_{L_B}M_{L_C}}(\boldsymbol{p}),
\end{eqnarray}
where $\langle \phi_B^{124} \phi_C^{35}|\phi_A^{123}\phi_0^{45}\rangle$ is the overlap of the flavor wave functions. The $I^{M_{L_A}m}_{M_{L_B}M_{L_C}}(\boldsymbol{p})$ are the spatial overlaps of the initial and final states, which can be written as
\begin{eqnarray}
I^{M_{L_A}m}_{M_{L_B}M_{L_C}}(\boldsymbol{p}) & = & \int d^3\boldsymbol{p}_1d^3\boldsymbol{p}_2d^3\boldsymbol{p}_3d^3\boldsymbol{p}_4d^3\boldsymbol{p}_5  \nonumber\\ && \times \delta^3(\boldsymbol{p}_1+\boldsymbol{p}_2+\boldsymbol{p}_3-\boldsymbol{P}_A)\delta^3(\boldsymbol{p}_4+\boldsymbol{p}_5)\nonumber\\ && \times \delta^3(\boldsymbol{p}_1+\boldsymbol{p}_4+\boldsymbol{p}_2-\boldsymbol{P}_B)\delta^3(\boldsymbol{p}_3+\boldsymbol{p}_5-\boldsymbol{P}_C) \nonumber\\
&& \times \psi^*_B(\boldsymbol{p}_1,\boldsymbol{p}_4,\boldsymbol{p}_2) \psi^*_C(\boldsymbol{p}_3,\boldsymbol{p}_5)\nonumber\\
&& \times\psi_A(\boldsymbol{p}_1,\boldsymbol{p}_2,\boldsymbol{p}_3){\cal{Y}}^m_1\left(\frac{\boldsymbol{p}_4-\boldsymbol{p}_5}{2}\right
).
\end{eqnarray}
It should be mentioned that the spatial overlap also relies on the radial and orbital quantum numbers that are commonly omitted for simplicity. The relevant spatial wave functions and overlaps are presented in Appendix.

In this work, we adopt the simplest vertex which assumes a spatially constant quark pair creation strength $\gamma$, the relativistic phase space, and the simple harmonic oscillator wave functions. Then, the decay width $\Gamma(A\rightarrow BC)$ can be calculated directly 
\begin{eqnarray}
\Gamma= \pi^2\frac{p}{M^2_A}\frac{1}{2J_A+1}\sum_{M_{J_A},M_{J_B},M_{J_C}}|{\cal{M}}^{M_{J_A}M_{J_B}M_{J_C}}|^2,
\end{eqnarray}
where $p=|\boldsymbol{p}|=\frac{\sqrt{[M^2_A-(M_B+M_C)^2][M^2_A-(M_B-M_C)^2]}}{2M_A}$,
and $M_A$, $M_B$, and $M_C$ are the masses of the hadrons $A$, $B$, and $C$, respectively.

Because there is no experimental information on doubly bottom baryons, we can not obtain the overall $\gamma$ and harmonic oscillator parameters by fitting the known decay processes. In Ref.~\cite{Xiao:2017dly}, the authors did not use the similarity between doubly charmed baryons and charmed mesons, then their parameters came from the singly heavy baryons. Here, owing to the rather heavy mass of bottom quark, the doubly bottom baryons are similar with the conventional bottom/bottom-strange mesons. Then, we prefer to adopt the same parameters of mesons to estimate the strong decays of doubly bottom baryons. Certainly, one can persist in calculating the strong decays for doubly bottom baryons with the same parameters taken from Ref.~\cite{Xiao:2017dly} straightforward, and will find that the main decay behaviors calculated by different parameters are self-consistent. In the literature~\cite{3p0model5,3p0model6}, the $\gamma = 0.4$ and $\alpha= 0.4~\rm{GeV}$ are widely used to investigate the strong decays of conventional mesons, where a factor of $\sqrt{96 \pi}$ should be added here by considering different filed conventions and the results of decay widths are unaffected certainly. Hence, the $\lambda$-mode harmonic oscillator parameter $\alpha_\lambda$ can be chosen as 0.4$~\rm{GeV}$, and the strong decay widths of low-lying states are independent with the $\rho$-mode harmonic oscillator parameter $\alpha_\rho$ that is absent in the overlaps of spatial wave functions.

Theoretically, we can also adopt the potential model to solve the mass spectrum, and then employ the realistic wave functions to calculate the decay behaviors. However, this approach needs to introduce more parameters. Here, we hope to use as few parameters as possible to study the strong decays of doubly bottom baryons semi-quantitatively, pick out narrow excited states and decay modes, and provide theoretical references for future experimental searches. Precise calculations require more experimental information in the future.
\section{STRONG DECAYS}{\label{low-lying}}
\subsection{$\Xi_{bb}(1P)$ states}
In the constituent quark model, there are five $\lambda$-mode $\Xi_{bb}(1P)$ states, and their predicted masses are about $10632 \sim 10694~\rm{MeV}$. According to approximately conserved light quark spin $j$, they belong to two groups: $j=1/2$ doublet and $j=3/2$ triplet. Their strong decay behaviors are shown in Table~\ref{1P1}. For the two $j=1/2$ states, the total decay widths are about 200 MeV, and the slight difference of total widths arises from the small mass splitting. The $\Xi_{bb} \pi$ and $\Xi_{bb}^{*} \pi$ decay channel saturate the total widths for $\Xi_{bb}( \frac{1}{2}^-, \frac{1}{2})$ and $\Xi_{bb} ( \frac{3}{2}^-,\frac{1}{2})$ states, respectively, which provides a good criterion to distinguish them. 

Unlike the broad $j=1/2$ states, the three $j=3/2$ states are relatively narrow. The strong decay of $\Xi_{bb} (\frac{1}{2}^-, \frac{3}{2} )$ state is governed by the $\Xi_{bb}^{*} \pi$ channel with a width of 27 MeV. Meanwhile, the total decay widths of $\Xi_{bb} (\frac{3}{2}^-,\frac{3}{2})$ and $\Xi_{bb} (\frac{5}{2}^-, \frac{3}{2}) $ states are about 37 and 53 MeV, respectively. The branching ratios for $J^P=3/2^-$ and $5/2^-$ states are predicted to be 
\begin{equation}
	Br(\Xi_{bb} \pi, \Xi_{bb}^{*} \pi) =28.4\%,71.6\%,
\end{equation}
and
\begin{equation}
	Br(\Xi_{bb} \pi, \Xi_{bb}^{*} \pi) =64.0\%,36.0\%.
\end{equation}
These ratios are independent with the overall quark pair creation strength $\gamma$ and can be tested in future. 

It can be seen that the total decay widths are quite different for different light quark spins $j$, and the partial decay widths are crucial to determine the spin-parities in a certain multiplet. Also, one can find the similarity between $\Xi_{bb}(1P)$ and $B(1P)$. Indeed, the $P$-wave bottom mesons can be categorized into two groups: one belongs to the $j=1/2$ doublet, and the other is the $j=3/2$ doublet. The two narrow $j=3/2$ bottom mesons have been observed experimentally, while the two $j=1/2$ states are predicted to be broad if they lie above the relevant threshold~\cite{Lu:2016bbk}. 

In the Ref~\cite{Eakins:2012fq}, the authors calculated the strong decays for these $P$-wave states within the potential model, and gives larger total decay widths for the $j=1/2$ doublet and $j=3/2$ triplet. Although the exact values for these decays are model dependent, their features are similar to ours. Also, the strong decays for $\Xi_{bb}(1P)$ states are investigated in the $L-S$ coupling scheme within the chiral quark model~\cite{Xiao:2017udy}, which can not be compared with present results in $j-j$ coupling scheme directly. Certainly, if the same coupling scheme is employed in these two works, the formulas of amplitudes between bottom and charmed sector should be same. However, the momenta of final states are also different in bottom and charmed sectors, which can affect the final branching ratios through the momentum-dependent amplitudes.

\begin{table}[!htbp]
	\begin{center}
		\caption{ \label{1P1} Theoretical predictions of the strong decays for the $\Xi_{bb}(1P)$ states in MeV.}
		\renewcommand{\arraystretch}{1.5}
		\footnotesize
		\begin{tabular*}{8.8cm}{@{\extracolsep{\fill}}p{1.55cm}<{\centering}p{1.3cm}<{\centering}p{1.3cm}<{\centering}p{1.3cm}<{\centering}p{1.3cm}<{\centering}p{1.45cm}<{\centering}}
			\hline\hline
			State  &   $\Xi_{bb} (\frac{1}{2}^-, \frac{1}{2})$ & $\Xi_{bb} (\frac{3}{2}^-, \frac{1}{2})$ & $\Xi_{bb} (\frac{1}{2}^-, \frac{3}{2})$	&			$\Xi_{bb} (\frac{3}{2}^-, \frac{3}{2})$ &   $\Xi_{bb} (\frac{5}{2}^-, \frac{3}{2}) $ \\
			\hline
			$\Xi_{bb} \pi$	  &		195.97	             &	$\cdots$	& $\cdots$	&	10.75	&34.01\\
			$\Xi_{bb}^{*} \pi$ &		$\cdots$  	&	205.64          &27.21             &27.06 	&19.14\\
			Total width    &        195.97                    &205.64                  &27.21              &37.81 &53.15\\
			\hline\hline
		\end{tabular*}
	\end{center}
\end{table}

\subsection{$\Xi_{bb}(2S)$ states}
In the $\Xi_{bb}$ family, there are two $\lambda$-mode first radially excited states, which can be denoted as $\Xi_{bb}(2S)$ and $\Xi_{bb}^{*}(2S)$. The total decay widths of $\Xi_{bb}(2S)$ and $\Xi_{bb}^{*}(2S)$ states are about 93 and 79 MeV, respectively, which are shown in Table~\ref{2S1}. Clearly, the masses, total widths, and main decay modes of these two states are similar, but the partial decay widths are different. The branching ratios are predicted to be 
\begin{equation}
	Br(\Xi_{bb} \pi, \Xi_{bb}^{*} \pi) =9.9\%,90.1\%.
\end{equation}
for the $\Xi_{bb} (2S)$ state, and 
\begin{equation}
	Br(\Xi_{bb} \pi, \Xi_{bb}^{*} \pi, \Omega_{bb} K) =39.4\%, 59.8\%, 0.8\%.
\end{equation}
for the $\Xi_{bb}^{*}(2S)$ state, which can help us to distinguish these two states. Hopefully, future experiments can search for them in the $\Xi_{bb} \pi$ and $\Xi_{bb}^* \pi$ final states.

Our results are quite different with the previous work in Ref.~\cite{Eakins:2012fq}, where significantly small total widths are predicted. The different initial masses and shapes of spatial wave functions could lead to this divergence. Actually, the estimated widths may be sensitive to the node of radially excited wave functions, which also appeared in previous $^3P_0$ model calculations~\cite{LeYaouanc:1977gm}. Moreover, the relativistic corrections may be important and provide significant contributions for these radially excited baryons~\cite{Arifi:2021orx}.       

\begin{table}[!htbp]
	\begin{center}
		\caption{ \label{2S1} Theoretical predictions of the strong decays for the $\Xi_{bb} (2S)$ and $\Xi_{bb}^{*}(2S)$ states in MeV.}
		\renewcommand{\arraystretch}{1.5}
		\footnotesize
		\begin{tabular*}{8.8cm}{@{\extracolsep{\fill}}*{3}{p{1.6cm}<{\centering}}}
			\hline\hline
			State  &   $\Xi_{bb} (2S)$ &  $\Xi_{bb}^{*}(2S)$		 \\
			\hline
			$\Xi_{bb} \pi$	  &		9.16	&	31.16	 \\
			$\Xi_{bb}^{*} \pi$ &		83.59	&	47.29	 \\
			$\Omega_{bb} K$   &           $\cdots$      &0.63   \\
			Total width    &        92.75      &79.08 \\
			\hline\hline
		\end{tabular*}
	\end{center}
\end{table}

\subsection{$\Xi_{bb}(1D)$ states}
In the constituent quark model, there are six states belonging to the $\lambda$-mode $\Xi_{bb}(1D)$ states, and they can be classified into $j=3/2$ triplet and $j=5/2$ triplet. These states lie in the range of  $11016 \sim 11086$ MeV, which are estimated by the Regge trajectory and heavy diquark symmetry. With this mass range, their strong decay behaviors are calculated and listed in Table~\ref{1D1}.  

It can be seen that the strong decay widths for these states are large, especially for the $j=5/2$ triplet. In Ref.~\cite{Eakins:2012fq}, only the pion emission processes were considered and falsely narrow states were obtained. From our results, the $K$, $\rho$, and $\omega$ meson emissions may be also important. Experimentally, the broad states can be hardly observed. The total decay widths of $\Xi_{bb} ( \frac{5}{2}^+, \frac{3}{2})$ state is about $130 \sim 156$ MeV, which is relatively narrower than others. Its main decay modes are $\Xi_{bb}^*\pi$ and $\Omega_{bb}^*K$, which can be tested by future experiments.

\begin{table*}[!htbp]
	\begin{center}
		\caption{ \label{1D1} Theoretical predictions of the strong decays for the $\Xi_{bb}(1D)$ states in MeV.}
		\renewcommand{\arraystretch}{1.5}
		\footnotesize
		\begin{tabular*}{18cm}{@{\extracolsep{\fill}}p{2cm}<{\centering}p{2.3cm}<{\centering}p{2.3cm}<{\centering}p{3cm}<{\centering}p{3cm}<{\centering}p{2.3cm}<{\centering}p{2.6cm}<{\centering}}
			\hline\hline
			State  &  $\Xi_{bb} (\frac{1}{2}^+, \frac{3}{2})$ & $\Xi_{bb} (\frac{3}{2}^+, \frac{3}{2})$	&$\Xi_{bb} (\frac{5}{2}^+, \frac{3}{2})$ &	$\Xi_{bb} (\frac{3}{2}^+, \frac{5}{2})$ &		 $\Xi_{bb} (\frac{5}{2}^+, \frac{5}{2}) $& $\Xi_{bb} (\frac{7}{2}^+, \frac{5}{2}) $\\
			\hline
			$\Xi_{bb} \pi$	      &  15.97-42.60   &	9.98-26.62     & $\cdots$    & $\cdots$ 	& 61.78-93.18       &148.27-223.63\\
			$\Xi_{bb}^{*} \pi$    & 3.54-7.16	   &	14.18-28.64    &31.92-64.44                   &205.29-321.96 &156.42-245.31     &87.98-137.99\\
			$\Omega_{bb} K$       &67.58 -84.47     & 42.24-52.80     & $\cdots$     & $\cdots$ & 1.24-4.66     & 2.97-11.17\\
			$\Omega_{bb}^{*} K$   &  6.92-9.94     & 27.69-39.76       &62.31-89.47              &  2.44-11.75                          &1.86-8.95      &1.05-5.04\\
			$\Xi_{bb}   \rho$     & 20.71-64.50    &13.99      -47.29 &2.80-18.61                 &59.67-186.22                      &34.88-111.32    &0.18-6.47\\
			$\Xi_{bb}^{*} \rho$   &0.77-44.76      &0.51-31.06         &0.09-8.24               &2.21-129.04                              &1.28-75.99     &0-1.71\\
			$\Xi_{bb}   \omega$   &5.31 -20.56     &3.58-14.88        &0.69-5.41                &15.29-59.33                               &8.93-35.34    &0.03-1.75\\
			$\Xi_{bb}^{*} \omega$ &0-13.50     &0-9.30       &0-2.30                 &0-38.90                                 &0-22.86   &0-0.41\\
			Total width           & 151.05-257.24  &143.27-219.25       &130.33-155.95                      &284.90-747.20                  &266.39-597.61     &240.48-388.17\\
			\hline\hline
		\end{tabular*}
	\end{center}
\end{table*}

\subsection{$\Omega_{bb}(1P)$ states}

In the quark model, there are five $\lambda$-mode $\Omega_{bb}(1P)$ states, which are named as $\Omega_{bb} (\frac{1}{2}^-, \frac{1}{2})$, $\Omega_{bb} (\frac{3}{2}^-, \frac{1}{2})$, $\Omega_{bb} (\frac{1}{2}^-, \frac{3}{2})$, $\Omega_{bb} (\frac{3}{2}^-, \frac{3}{2})$ and  $\Omega_{bb} (\frac{5}{2}^-, \frac{3}{2})$, respectively. The estimated masses for these $\Omega_{bb}(1P)$ states are around 10800 MeV. With these initial masses, the total widths of two $j=1/2$ $\Omega_{bb}$ states are rather broad, which can not be observed experimentally. For the three $j=3/2$ states, the total widths are narrow, which have good potentials to be observed in the $\Xi_{bb} \bar K$ and $\Xi_{bb}^* \bar K$ modes.

From the heavy diquark symmetry, these $\Omega_{bb}(1P)$ states can be related to the $B_s(1P)$ mesons, where the broad $j=1/2$ $\Omega_{bb}(1P)$ doublet correspond to the $j=1/2$ $B_s(1P)$ states and  narrow $j=3/2$ $\Omega_{bb}(1P)$ triplet correspond to the $j=3/2$ $B_s(1P)$ states. Also, the two narrow $j=3/2$ $B_s(1P)$ states, $B_{s1}(5830)$ and $B_{s2}^*(5840)$, have been observed experimentally~\cite{pdg}, which supports our calculations. Meanwhile, the two predicted $j=1/2$ $B_s(1P)$ states are not found until now.

However, if we go further to consider the heavy quark flavor symmetry, the two $j=1/2$ $B_s(1P)$ states may have similar properties with $D_{s0}^*(2317)$ and $D_{s1}(2460)$. Then, the $j=1/2$ $\Omega_{bb}(1P)$ doublet are also related with $D_{s0}^*(2317)$ and $D_{s1}(2460)$ resonances. In this situation, these two $j=1/2$ $\Omega_{bb}(1P)$ states should lie below the $\Xi_{bb} \bar K$ and $\Xi_{bb}^* \bar K$ thresholds, and the isospin broken and radiative modes may dominate the decay behaviors. Future experiments can help us to disentangle this puzzle and deepen our understandings of the mysterious $D_{s0}^*(2317)$ state. 

\begin{table}[!htbp]
	\begin{center}
		\caption{ \label{1P} Theoretical predictions of the strong decays for the  $\Omega_{bb}(1P)$ states in MeV.}
		\renewcommand{\arraystretch}{1.5}
		\footnotesize
		\begin{tabular*}{8.8cm}{@{\extracolsep{\fill}}p{1.55cm}<{\centering}p{1.3cm}<{\centering}p{1.3cm}<{\centering}p{1.3cm}<{\centering}p{1.3cm}<{\centering}p{1.4cm}<{\centering}}
			\hline\hline
			Mode  &  $\Omega_{bb} (\frac{1}{2}^-, \frac{1}{2})$ &$\Omega_{bb} (\frac{3}{2}^-, \frac{1}{2})$&$\Omega_{bb} (\frac{1}{2}^-, \frac{3}{2})$  &$\Omega_{bb} (\frac{3}{2}^-, \frac{3}{2})$&$\Omega_{bb} (\frac{5}{2}^-, \frac{3}{2}) $\\
			\hline
			$\Xi_{bb} \bar K$	           &		485.99	   &$\cdots$         &$\cdots$	           & 4.39 &16.79 \\
			$\Xi_{bb}^{*} \bar K$	      &		$\cdots$	 &492.92                & 2.62	           &4.66   &4.82 \\   
			Total width                  &     485.99                  &  492.92          &  2.62                 &   9.05 &  21.61 \\
			\hline\hline
		\end{tabular*}
	\end{center}
\end{table}

\subsection{$\Omega_{bb}(2S)$ states}

In the quark model, there are two $\lambda$-mode $2S$ states, which can be denoted as $\Omega_{bb}(2S)$ and $\Omega_{bb}^*(2S)$. The total decay widths of $\Omega_{bb}(2S)$ and $\Omega_{bb}^*(2S)$ states are predicted to be 184 and 175 MeV, respectively, which are listed in Table~\ref{2S2}. These two states have similar masses, total widths, and dominant decay modes, but the branching ratios are different. From Table~\ref{2S2}, the predicted branching ratios are  
\begin{equation}
Br(\Xi_{bb} \bar K, \Xi_{bb}^{*} \bar K, \Omega_{bb} \eta, \Omega_{bb}^{*} \eta) =9.5\%, 85.5\%, 1.1\%, 3.9\%.
\end{equation}
for the $\Omega_{bb}(2S)$ state, and 
\begin{equation}
Br(\Xi_{bb} \bar K, \Xi_{bb}^{*} \bar K, \Omega_{bb} \eta, \Omega_{bb}^{*} \eta) =35.5\%, 52.7\%, 6.8\%, 5.0\%.
\end{equation}
for the $\Omega_{bb}^{*}(2S)$ state. The branching ratios of $\Xi_{bb} \bar K$ and $\Xi_{bb}^{*} \bar K$ channels are significantly different, which can help us to distinguish $\Omega_{bb}(2S)$ and $\Omega_{bb}^{*}(2S)$ states in future.

\begin{table}[!htbp]
\begin{center}
\caption{ \label{2S2} Theoretical predictions of the strong decays for the $\Omega_{bb}(2S)$ and $\Omega_{bb}^{*}(2S)$ states in MeV.}
\renewcommand{\arraystretch}{1.5}
\footnotesize
\begin{tabular*}{8.8cm}{@{\extracolsep{\fill}}*{3}{p{1.6cm}<{\centering}}}
\hline\hline
Mode  &   $\Omega_{bb}(2S)$ & $\Omega_{bb}^{*}(2S)$  \\
\hline
$\Xi_{bb} \bar K$	           &		17.55	 &62.24   \\
$\Xi_{bb}^{*} \bar K$	      &		157.26	&	92.49	 \\   
$\Omega_{bb} \eta$          &      2.08   &  11.89         \\
$\Omega_{bb}^{*} \eta $     &      7.10 &   8.79        \\
Total width                  &      183.99  &  175.41          \\
\hline\hline
\end{tabular*}
\end{center}
\end{table}

\subsection{$\Omega_{bb}(1D)$ states}

The strong decays for the $\lambda$-mode $\Omega_{bb}(1D)$ states are listed in Table~\ref{1D}. It can be found that all these states have large total widths, especially for the $j=5/2$ triplet, which can be hardly observed in experiments. Owing to the higher initial masses, the $\Omega_{bb}(1D)$ states may also decay into the $\Xi_b \bar B$ final states. This singly heavy baryon plus heavy meson decay mode may be important for the higher excited states but can be neglected for the low-lying states due to limited phase space. With the light quark SU(3) flavor symmetry, these $\Omega_{bb}(1D)$ states should have similar decay behaviors with $\Xi_{bb}(1D)$ states, which agrees with our present calculations.   

\begin{table*}[!htbp]
\begin{center}
\caption{ \label{1D} Theoretical predictions of the strong decays for the  $\Omega_{bb}(1D)$ states in MeV.}
\renewcommand{\arraystretch}{1.5}
\footnotesize
\begin{tabular*}{18cm}{@{\extracolsep{\fill}}p{2cm}<{\centering}p{2.3cm}<{\centering}p{2.3cm}<{\centering}p{3cm}<{\centering}p{3cm}<{\centering}p{2.3cm}<{\centering}p{2.6cm}<{\centering}}
\hline\hline
Mode  &  $\Omega_{bb} (\frac{1}{2}^+, \frac{3}{2})$ & $\Omega_{bb} (\frac{3}{2}^+, \frac{3}{2})$ &$\Omega_{bb} (\frac{5}{2}^+, \frac{3}{2})$ &$\Omega_{bb} (\frac{3}{2}^+, \frac{5}{2})$&$\Omega_{bb} (\frac{5}{2}^+, \frac{5}{2}) $&$\Omega_{bb} (\frac{7}{2}^+, \frac{5}{2}) $\\
\hline 
$\Xi_{bb} \bar K$	           &29.47-46.48	   &1.43-29.05	     &$\cdots$    & $\cdots$ &113.47-133.38       &272.31-320.11 \\
$\Xi_{bb}^{*} \bar K$	      &6.61-9.08         & 26.42-36.32	&59.48-81.71	                &374.95-449.56                             &285.68-342.52      & 160.69-192.67\\   
$\Omega_{bb} \eta$          &41.73-44.03        &26.08-27.52      &$\cdots$      &$\cdots$ &5.14-7.10         &12.33-17.03\\ 
$\Omega_{bb}^{*} \eta $     &   5.52-5.53     & 22.14-22.47     &49.81-50.56                     &14.19-20.48                             &10.81-15.61       & 6.08-8.78 \\
$\Xi_{bb}  K^{*}$           &42.15-73.28       &65.44-98.85      &104.25-141.47                    &65.17-93.78                            &158.55-209.78      &289.29-372.18\\
$\Xi_{bb}^{*}  K^{*}$       &16.2234.09       &29.39-55.40      &51.34-90.93                     &30.59-55.96                             &79.61-139.18      &148.24-255.70\\
Total width                  &  172.29-193.00    &215.63-248.74   &  287.86-341.69                  & 484.90-619.78                     &  661.85-861.53    &909.54-1199.97\\
\hline\hline
\end{tabular*}
\end{center}
\end{table*}

\subsection{Low-lying $\rho$-mode and $\rho$-$\lambda$ hybrid  states}

Together with the $\lambda$-mode excited states, there also exist lots of $\rho$-mode and $\rho$-$\lambda$ hybrid excited states. For the doubly heavy baryons, the $\rho$-mode between two heavy quarks is more easily excited due to its larger reduced masses. Then, the masses of low-lying $\rho$-mode and $\rho$-$\lambda$ hybrid doubly bottom states with $N \leq 2$ should be smaller than that of $\lambda$-mode $\Xi_{bb}(1D)$ or $\Omega_{bb}(1D)$ states. One can see that these low-lying $\rho$-mode and $\rho$-$\lambda$ hybrid states should lie below the $\Lambda_b \bar B$  or $\Xi_b \bar B$ threshold, and the singly heavy baryon plus heavy meson decay modes are forbidden due to the insufficient phase space. 

The other possible decay modes for these low-lying $\rho$-mode and $\rho$-$\lambda$ hybrid states are the light meson emission processes as well as the $\lambda$-mode states. However, under spectator assumption of $^3P_0$ model, the spatial wave functions between initial $\rho$-mode and $\rho$-$\lambda$ hybrid doubly bottom baryons and final ground states are orthogonal, which leads to the vanishing amplitudes and strong decay widths. It can be seen that the $^3P_0$ model preserves the heavy diquark symmetry automatically, where the heavy quark subsystems with different quantum numbers can not transit into each other through the transition operator $T$. Then, these states should be rather narrow, and the radiative and weak decays become crucial, which provides good opportunities to be searched by future experiments.

\section{SUMMARY}
In this work, we investigate the strong decays of low-lying doubly bottom baryons within the $^3P_0$ model systematically. The relevant formulas of $^3P_0$ model are constructed in the $j-j$ coupling scheme, where the heavy diquark symmetry are preserved. The masses of $S$- and $P$-wave doubly bottom states are taken form the relativistic quark model, and the average masses of $\lambda$-mode $\Xi_{bb}(1D)$ and $\Omega_{bb}(1D)$ states are estimated with the help of Regge trajectory and heavy diquark symmetry. Then, the strong decays of these low-lying doubly bottom baryons are calculated.     

Our results show that some of $\lambda$-mode $\Xi_{bb}(1P)$ and $\Omega_{bb}(1P)$ states are rather narrow, which have good potentials to be observed by future experiments. The other $\lambda$-mode states are relatively broad, which makes it difficult to search for. Moreover, the narrow excited states may be observed more easily than the ground states in future. For instance, in the singly bottom baryons, the four excited $\Omega_{b}(1P)$ states were observed by LHCb Collaboration recently~\cite{LHCb:2020tqd}, while one of the ground states, $\Omega_{b}^{*}$, is missing till now. For the low-lying $\rho$-mode and $\rho$-$\lambda$ hybrid states, the Okubo-Zweig-Iizuka-allowed strong decays are highly suppressed owing to the orthogonality of spatial wave functions between initial and final doubly bottom baryons. These $\rho$-mode and $\rho$-$\lambda$ hybrid states should be extremely narrow and the radiative and weak decays become crucial, and future experiments can test our phenomenological predictions at the quark level.   

In the heavy quark limit, the two heavy quark subsystem in a doubly heavy baryon seems like a antiquark and a heavy diquark symmetry emerges. Actually, it can be noticed that the $^3P_0$ model is  a spectator model, where the quarks in the initial state carry their color, flavor, spin, and momenta into the final states, and the change for degrees of freedom arises from the created quark pair. Under this hypothesis, the two heavy quarks as a whole go into the final states, and the heavy diquark symmetry is preserved automatically.  

\begin{appendix}
\section*{Appendix A Overlaps of spatial wave functions}
 \label{A}
\addcontentsline{toc}{chapter}{Appendix A Overlaps of spatial wave functions }

\setcounter{equation}{0}
\renewcommand\theequation{A.\arabic{equation}}

The harmonic oscillator wave functions for doubly bottom baryons in momentum representation can be expressed as 
\begin{eqnarray}
&&\psi \left( n_{\rho },l_{\rho },m_{\rho },n_{\lambda },l_{\lambda },m_{\lambda }\right)  =  \nonumber \\  && \left(
-1\right) ^{n_{\rho }} \left(
-i\right) ^{l_{\rho }}{p_{\rho}^{l_{\rho}}} {\left[ \frac{2n_{\rho} !}{ \Gamma \left (
n_{\rho}+l_{\rho }+3/2 \right) }\right]} ^{\frac{1}{2}}
 \left( \frac{1}{\alpha _{\rho }} \right) ^{\frac{3}{2}+l_{\rho }} L_{n_{\rho}}^{{l_\rho}+ \frac{1}{2}}({p_{\rho}^2}/{\alpha _{\rho }^2})\nonumber \\
&& \times
 \left(
-1\right) ^{n_{\lambda }}\left(
-i\right) ^{l_{\lambda }}  {p_{\lambda}^{l_{\lambda }}} {\left[ \frac{2n_\lambda !}{ \Gamma \left(
n_\lambda+l_{\lambda }+3/2\right) }\right]} ^{\frac{1}{2}}
 \left( \frac{1}{\alpha _{\lambda }} \right) ^{\frac{3}{2}+l_{\lambda} } L_{n_\lambda}^{{l_\lambda}+ \frac{1}{2}}({p_{\lambda}^2}/{\alpha _{\lambda }^2}) 
 \nonumber  \\
&&\times \exp \left( -\frac{\vec{p}_{\rho }^{2}}{2\alpha _{\rho }^{2}}-\frac{\vec{p}_{\lambda }^{2}}{2\alpha _{\lambda }^{2}}\right)  Y_{l_{\rho }}^{m_{\rho }}\left( \vec{p}_{\rho }\right) Y_{l_{\lambda }}^{m_{\lambda }}\left( \vec{p}_{\lambda }\right) ,
\end{eqnarray}%
where $\vec{p}_{\rho }=\frac{1}{\sqrt{2}}\left( \vec{p}_{1}-\vec{p}_{2}\right)$, $\vec{p}_{\lambda }=\frac{1}{\sqrt{6}}\left( \vec{p}_{1}+\vec{p}_{2}-2\vec{p}%
_{3}\right)$, and $Y_{l}^{m}(\vec{p})$ is a three-dimensional spherical harmonic function. Similarly, the harmonic oscillator wave function for ground mesons in momentum representation can be written as    
\begin{eqnarray}
\psi \left(0,0,0\right)=\left( \frac{1}{\pi \alpha^{2}}\right) ^{\frac{3}{4}}\exp \left( -%
\frac{\vec{p}_{rel}^{2}}{2\alpha^{2}}\right),
\end{eqnarray}%
where $\vec{p}_{rel}$ represents the relative momentum between the quark and antiquark in the final mesons. 

In this paper, all the final states are ground states, that is, $n_{\rho_{B}}$=$l_{\rho_{B}}$=$n_{\lambda_{B}}$=$l_{\lambda_{B}}$=$L_{C}$=0. Here, we denote the spatial overlap integrals $I^{M_{L_A},m}_{M_{L_B},M_{L_C}}(\vec{p})$ as $\Pi(n_{\rho_{A}},l_{\rho_{A}},m_{\rho_{A}}, n_{\lambda_{A}},l_{\lambda_{A}},m_{\lambda_{A}},m)$, and the relevant formulas for the low-lying states are present as follows.

Define 
\begin{eqnarray}
f _{1}=\frac{1}{2\alpha _{\lambda}^{2}}+\frac{1}{2\alpha _{\lambda}^{^{\prime }2}}+\frac{1}{3 \alpha^{2}},
\end{eqnarray}%
\begin{eqnarray}
f _{2}=\frac{2}{\sqrt{6}\alpha _{\lambda
}^{^{\prime }2}}+\frac{1}{\sqrt{6} \alpha^{2}},
\end{eqnarray}%
\begin{eqnarray}
f _{3}=\frac{1}{3\alpha _{\lambda}^{^{\prime }2}}+
\frac{1}{8 \alpha^{2}},
\end{eqnarray}%
\begin{eqnarray}
\beta =1-\frac{f _{2}}{\sqrt{6}f _{1}},
\end{eqnarray}%
and then we can obtain the spatial overlaps integrals straightforwardly
\begin{eqnarray}
\Pi \left( 0,0,0,0,0,0,0\right) =\beta \left\vert \vec{p}\right\vert \Delta _{00},
\end{eqnarray}%
\begin{eqnarray}
\Pi \left( 0,0,0,1,0,0,0\right) &=&\Bigg (-\sqrt{\frac{3} {2}} \beta \left\vert \vec{p}\right\vert +\sqrt{\frac{3} {2}} \frac{\beta \left\vert \vec{p}\right\vert}{\alpha_\lambda^2 f_1}+\frac{f_2^2\beta \left\vert \vec{p}\right\vert^3}{2\sqrt{6}\alpha_\lambda^2 f_1^2} \nonumber \\ &&  -\frac{f_2 \left\vert \vec{p}\right\vert}{3 \alpha_\lambda^2 f_1^2} \Bigg )\Delta _{00},
\end{eqnarray}%
\begin{eqnarray}
\Pi \left( 0,0,0,0,1,0,0\right) =\left( \frac{1}{\sqrt{6} f _{1}}-\frac{f _{2}}{2f _{1}}\beta \left\vert \vec{p}\right\vert ^{2} \right)  \Delta _{01},
\end{eqnarray}%
\begin{eqnarray}
\Pi \left( 0,0,0,0,1,1,-1\right) &=&\Pi \left(
0,0,0,0,1,-1,1\right)  \nonumber \\ &=& -\frac{1}{\sqrt{6} f _{1}}\Delta _{01},
\end{eqnarray}%
\begin{eqnarray}
\Pi \left( 0,0,0,0,2,0,0\right) =\left( \frac{f _{2}^{2}}{2f _{1}^{2}}\beta
\left\vert \vec{p}\right\vert ^{3}- \frac{\sqrt{6} f _{2}}{3f _{1}^{2}} \left\vert \vec{p}\right\vert
\right) \Delta _{02},
\end{eqnarray}%
\begin{eqnarray}
\Pi \left( 0,0,0,0,2,1,-1\right) &=& \Pi \left( 0,0,0,0,2,-1,1\right) \nonumber \\ &=&  \frac{f _{2}}{\sqrt{2}f _{1}^{2}}
\left\vert \vec{p}\right\vert \Delta _{02}, 
\end{eqnarray}%
with  
\begin{eqnarray}
\Delta _{00} &=&\left( {\frac{\sqrt3}{2 \pi ^{\frac{5}{4}}}}\right)
\left( \frac{1}{\alpha _{\lambda }^{\prime } \alpha_{\lambda }f _{1}\alpha }\right) ^{%
\frac{3}{2}}
  \nonumber \\
&&\times \exp \left[ -\left( f _{3}-\frac{f _{2}^{2}}{4f _{1}}\right) \left\vert \vec{p%
}\right\vert ^{2}\right], 
\end{eqnarray}%
\begin{eqnarray}
\Delta _{01} &=&\left( i\sqrt{\frac{3}{2 \pi ^{\frac{5}{2}}}}\right)
\left( \frac{1}{f _{1}\alpha \alpha _{\lambda }^{\prime }}\right) ^{\frac{3}{2}}  
  \left( \frac{1}{\alpha _{\lambda }}\right) ^{\frac{5}{2}}  
\nonumber \\
&&\times \exp \left[ -\left(f _{3}-\frac{f _{2}^{2}}{4f _{1}}\right) \left\vert \vec{p%
}\right\vert ^{2}\right],
\end{eqnarray}%

\begin{eqnarray}
\Delta _{02} &=&\left( \frac{1}{2 \pi ^{\frac{5}{4}}}\right)
\left( \frac{1}{f _{1}\alpha \alpha _{\lambda }^{\prime }}\right) ^{\frac{3}{2}} 
\left( \frac{1}{\alpha _{\lambda }}\right) ^{\frac{7}{2}} 
 \nonumber \\
&&\times \exp \left[ -\left(f _{3}-\frac{f _{2}^{2}}{4f _{1}}\right) \left\vert \vec{p%
}\right\vert ^{2}\right].
\end{eqnarray}%

\end{appendix}

\bigskip
\noindent
\begin{center}
	{\bf ACKNOWLEDGEMENTS}\\
	
\end{center}
This work is supported by the National Natural Science Foundation of China under Grants No. 11705056 and No. U1832173, and by the State Scholarship Fund of China Scholarship Council under Grant No. 202006725011.

\end{document}